\documentclass[12pt,russian]{article}
 \usepackage{helvet}
 \usepackage[T1]{fontenc}
 \usepackage[cp1251]{inputenc}
 \usepackage{geometry}
 \usepackage{graphicx}
 \usepackage{setspace}
 \usepackage{babel}

 \renewcommand{\vec}[1]{\mbox{\boldmath $#1$}}

 \def\n{${\cal N}^{\underline\circ}$}
 \def\rot{\mathop{\rm rot}\nolimits}

 \def\gsim{\lower.4ex\hbox{$\;\buildrel >\over{\scriptstyle\sim}\;$}}
 \def\lsim{\lower.4ex\hbox{$\;\buildrel <\over{\scriptstyle\sim}\;$}}
 \def\newpage{\vfill\eject}
 \def\bl{\par\vskip 12pt\noindent}
 \def\bll{\par\vskip 24pt\noindent}

 \def\n{${\cal N}^{\underline\circ}$}

 \def\aa{A\&A}
 \def\apj{ApJ}
 \def\apjs{ApJS}
 \def\an{Astron. Nachr.}

 \def\lrsp{Living Rev. Solar Phys.}

 \def\raa{Res. Astron. Astrophys.}
 \def\sp{Solar Phys.}
 \def\aj{Astron. Rep.}
 \def\paj{Astron. Lett.}

\begin{document}

\vskip 1.5 cm

\begin{center}
PARAMETRIC MODULATION OF DYNAMO WAVES
\end{center}

\bll

\centerline{L.\,L.~Kitchatinov$^{1,2}$\footnote{E-mail:kit@iszf.irk.ru}, A.\,A.~Nepomnyashchikh$^{1,3}$}

\bl

\begin{center}
$^1${\it Institute for Solar-Terrestrial Physics, Lermontov Str. 126A, Irkutsk, 664033, Russia} \\[0.1 truecm]
$^2${\it Pulkovo Astronomical Observatory, Pulkovskoe Sh. 65, St. Petersburg, 196140, Russia} \\[0.1 truecm]
$^3${\it Irkutsk State University, Gagarin Str. 20, 664003, Russia}
\end{center}

\bll
\hspace{0.8 truecm}
\parbox{14.4 truecm}{
Long-term variations of solar activity, including the Grand minima, are believed to result from temporal variations of dynamo parameters. The simplest approximation of dynamo waves is applied to show that cyclic variations of the parameters can lead to an exponential growth or decay of magnetic oscillations depending on the variations frequency. There is no parametric resonance in a dynamo, however: the selective sensitivity to distinct frequencies, characteristic of resonant phenomena, is absent. A qualitative explanation for this finding is suggested. Nonlinear analysis of dynamo-waves reveals the hysteresis phenomenon found earlier in more advanced models. However, the simplified model allows a computation of a sufficiently large number of dynamo-cycles for constructing the distribution function of their amplitudes to reproduce qualitatively two modes of solar activity inferred recently from cosmogenic isotope content in natural archives.
 }

\bll


\newpage

\reversemarginpar

\setlength{\baselineskip}{0.8 truecm}

 \centerline{INTRODUCTION}
 \bl

Hydrodynamic flows generating cosmic magnetic fields  usually vary with time. Para\-me\-ters of the generation processes - hydromagnetic dynamos - are, therefore, time-dependent. Variations of dynamo-parameters can be either periodic, as in the case of modulation by density waves in galactic dynamos, or irregular, as in stellar convection zones. The case of irregular variations currently attracts much attention in relation to the Grand minima of solar activity (cf., e.g., the review of Charbonneau 2010), events similar to the famous Maunder minimum when there were almost no sunspots for about 70 years (Hoyt \& Schatten 1996).

The number of large-scale convective cells generating magnetic fields in the Sun is not large (Gilman \& Miller 1985) so the longitude-averaged dynamo parameters fluctuate considerably (Olemskoy et al. 2013).  Fluctuations of dynamo-parameters are believed to be responsible for the observed variations of solar cycle durations and amplitudes and even for the events of Grand minima. Dynamo models with fluctuating parameters generally reproduce these observed phenomena (cf., e.g., Moss et al. 2008; Usoskin et al. 2009; Choudhuri \& Karak 2012; Karak \& Choudhuri 2013; Olemskoy \& Kitchatinov 2013). Relations between the fluctuation properties and the resulting variations in magnetic cycles remain, however, unknown. In particular, the important theoretical question on the possibility of parametric resonance in dynamos\footnote{The non-periodic character of the parametric fluctuations is significant but not of principal importance. Stochastic parametric resonance (Klyatskin 1980) should be considered for this case.} remains unanswered. It is not clear whether there are special frequencies in the fluctuations spectrum to which the field generation processes are especially sensitive. In some dynamo models, the characteristic fluctuation time was put equal to a half of the magnetic cycle period, which corresponds to the parametric resonance condition for harmonic oscillator, but an enhanced reaction on fluctuations of such a duration was not found. The reason probably is that contemporary dynamo models are numerical and it is difficult to find functional relations from the results of numerical computations.

Probably for the same reason, the publications on resonant phenomena in dynamos are not many. The dynamo equations do not possess sources of magnetic fields. The resonance problem, therefore, does not have an analogue of driving force, and only parametric resonance can be considered. Schmitt \& R\"udiger (1992) and Moss (1996) studied the parametric resonance for galactic dynamos. Gilman \& Dikpati (2011) and Moss \& Sokoloff (2013) considered the resonance phenomena for stellar convective dynamos. An interesting version of parametric resonance in nonlinear dynamos was found by Reshetnyak (2010): the resonance can result from a time-delay in the reaction of dynamo-parameters to the generated magnetic field.

This paper suggests using a simplified dynamo model to treat the resonance problem analytically. Analytical solutions may be helpful in understanding the results of more complicated numerical models. Such an approach proved to be productive in studies of magneto-rotational instability where a simplified analytical model revealed similarity rules confirmed subsequently by numerical computations (Kitchatinov \& R\"udiger 2004). Sufficient simplification for an analytical treatment of the resonance problem give an (local) approximation of dynamo-waves. In this approximation, the dynamo-wave length is assumed short compared to the scale of variations of dynamo parameters. Such a scale ratio does not hold for the Sun and for corresponding numerical models of global dynamos. However, it is difficult to find a numerical model whose results were not interpretable in terms of the Yoshimura (1975) law for dynamo-waves propagation.

We shall see that the parametric modulation of dynamo waves can lead to an exponential growth or decay of the wave amplitude depending on the frequency $\gamma$ of variation of the $\alpha$-parameter of magnetic field generation. This modulation, however, is not a resonance. The enhanced selective reaction to special frequencies, characteristic of resonant phenomena, is not present in dynamos. A qualitative explanation for this circumstance will be suggested.

Another problem where the dynamo-wave approximation can be useful is related to the two distinct modes of solar activity inferred recently by Usoskin et al. (2014) from radiocarbon data. The two modes may be related to the hysteresis in nonlinear dynamos: allowance for the dependence of turbulent diffusivity on the magnetic field reveals the coexistence of two stable dynamo solutions with different amplitudes of magnetic cycles that are realised depending on the pre-history of dynamo parameter variations (Kitchatinov \& Olemskoy 2010). Dynamo-hysteresis was also found in 3D numerical simulations of Karak et al. (2015). However, a moderate number of magnetic cycles covered by computations with complicated models hinders their  statistical analysis. We shall see that allowance for nonlinearities in the dynamo-wave approximation also shows the hysteresis phenomenon. Millions of magnetic cycles can be computed with this approximation. The distribution function of the amplitudes of these cycles has two maxima similar to the dual distribution found by Usoskin et al. (2014) from historical data on solar activity.
 \bll
 \centerline{BASIC EQUATIONS}
 \bl

We start from the mean-field dynamo equation
\begin{equation}
    \frac{\partial\vec{B}}{\partial t} = \rot\left( \alpha\vec{B}
        + \vec{V}\times\vec{B}
        -\eta\rot\vec{B}\right)
    \label{1}
\end{equation}
(cf., e.g., Zeldovich et al., 1983). The velocity $\vec{V}$ corresponds to the shear flow of, e.g., non-uniform rotation. Cartesian coordinates are used with the $y$-axis aligned with the velocity vector $\vec{V}$ and the $x$-axis along the velocity gradient,
\begin{equation}
    \vec{V} = \left( 0, Sx, 0\right) .
    \label{2}
\end{equation}
Coefficients in the equation (\ref{1}) and the shear value $S$ are assumed position-independent, i.e. the scale of magnetic field variation is assumed small compared to the characteristic scale of the fluid inhomogeneity  (short dynamo-waves). For the convection zone of a star, the $y$-axis of our coordinate system corresponds to the azimuthal direction, the $x$-axis points in the direction of the angular velocity gradient and the $yz$-plane is the isorotation surface. The magnetic field,
\begin{equation}
    \vec{B} = \left( -\frac{\partial A}{\partial z}, B, 0\right) ,
    \label{3}
\end{equation}
includes toroidal ($B$) and poloidal ($-\partial A/\partial z$) components.

The field $B$ and the poloidal field potential $A$ depend only on the coordinate $z$ and time. The equation system for the two field components reads
\begin{eqnarray}
    \frac{\partial B}{\partial t} &=&
    -S\frac{\partial A}{\partial z}
    + \eta\frac{\partial^2 B}{\partial z^2} ,
    \nonumber \\
    \frac{\partial A}{\partial t} &=& \alpha B + \eta\frac{\partial^2A}{\partial z^2},
    \label{4}
\end{eqnarray}
where the contribution of the $\alpha$-effect into the toroidal field generation is omitted (the $\alpha\Omega$-dynamo).

We consider wave-type solutions of $A,B \propto \mathrm{exp}(ikz)$ and normalize the equations (\ref{4}) to dimensionless units. Time is measured in units of $(k^2\eta_0)^{-1}$. The $\eta_0$ is the characteristic value of diffusivity, $\eta = \eta_0\hat\eta$ ($\hat\eta \sim 1$), and similar scaling is used for the $\alpha$-coefficient: $\alpha = \alpha_0\hat\alpha$. The potential $A$ is measured in units of $\alpha_0B_0/(\eta_0k^2)$; $B_0$ is the characteristic value of the field $B$ and simultaneously the unit for field measurement. We keep the same notations for normalised variables as used before for their dimensional counterparts to find
\begin{eqnarray}
    \dot{B} &=& -i{\cal D} A - \hat\eta B ,
    \nonumber \\
    \dot{A} &=& \hat\alpha B - \hat\eta A,
    \label{5}
\end{eqnarray}
where
\begin{equation}
    {\cal D} = \frac{\alpha_0 S}{\eta_0^2 k^3}
    \label{6}
\end{equation}
is the dynamo-number. The parameters $\hat\alpha$ and $\hat\eta$ may depend on time explicitly, as they do in the linear problem on parametric resonance, or implicitly via their dependence on the magnetic field amplitude in a nonlinear problem.
 \bll
 \centerline{ABSENCE OF PARAMETRIC RESONANCE}
 \bl

If the parameters of Eqs.\,(\ref{5}) do not depend on time ($\hat\eta = \hat\alpha = 1$), the threshold value of the dynamo-number for emergence of non-decaying oscillations ${\cal D} = 2$ (the amplitude of the oscillations grows exponentially with time for the larger dynamo number and decays with time for the smaller $D$). The eigenvalue problem ($A,B \propto \mathrm{exp}(-i\omega t)$) gives two solutions for this case,
\begin{equation}
    \omega_1 = 1,\ \ \ \omega_2 = -1 - 2i ,
    \label{7}
\end{equation}
the first of which corresponds to non-decaying oscillations. Let us consider how this solution changes with allowance for small variations of the $\hat\alpha$-parameter with time:
\begin{equation}
    \hat\alpha = 1 + \varepsilon a(t) ,
    \label{8}
\end{equation}
where $\varepsilon \ll 1$ is the amplitude of the parameter perturbations and $a \sim 1$. On eliminating $A$ from equations (\ref{5}), we find
\begin{equation}
    \ddot{B} + 2\dot{B} + \left( 1+ 2i + 2i\varepsilon a(t)\right) B = 0.
    \label{9}
\end{equation}

The problem of parametric resonance concerns periodic variations of the parameter with time: $a = \sin (\gamma t)$. We search for the solution in the form
\begin{equation}
    B = \mathrm{e}^{\sigma t}\sum\limits_{n=-\infty}^{\infty}b_n\mathrm{e}^{i(n\gamma-1)t} .
    \label{10}
\end{equation}
It can be seen that a decomposition of the coefficients $b_n$ in powers of the small parameter $\varepsilon$ starts with $\varepsilon^{\mid n\mid}$. With accuracy up to the second order in  $\varepsilon$, we find the amplitudes of \lq satelite' oscillations,
\begin{equation}
    b_1 = \frac{\varepsilon\ b_0}{\gamma(\gamma - 2i - 2)},\ \ \
    b_{-1} = -\frac{\varepsilon\ b_0}{\gamma(\gamma + 2i + 2)} ,
    \label{11}
\end{equation}
and the eigenvalue
\begin{equation}
    \sigma = s + i\Delta\omega,\ \
    s = \frac{\varepsilon^2}{2}\frac{\gamma^2 - 8}{\gamma^4 + 64},\ \
    \Delta\omega = \frac{\varepsilon^2}{2}\frac{\gamma^2 + 8}{\gamma^4 + 64},
    \label{12}
\end{equation}
including the (real) growth rate $s$ and frequency shift $\Delta\omega$. The approximate solution  (\ref{11}), (\ref{12}) is valid for the small ratio $\varepsilon/\gamma \ll 1$. It can be seen from the equation (\ref{12}) for $s$ that either exponential growth (for $\gamma > \sqrt{8}$) or decay (for $\gamma < \sqrt{8}$) of the oscillations is possible.

\begin{figure}[htb]
 \begin{center}
 \includegraphics[width=12 cm]{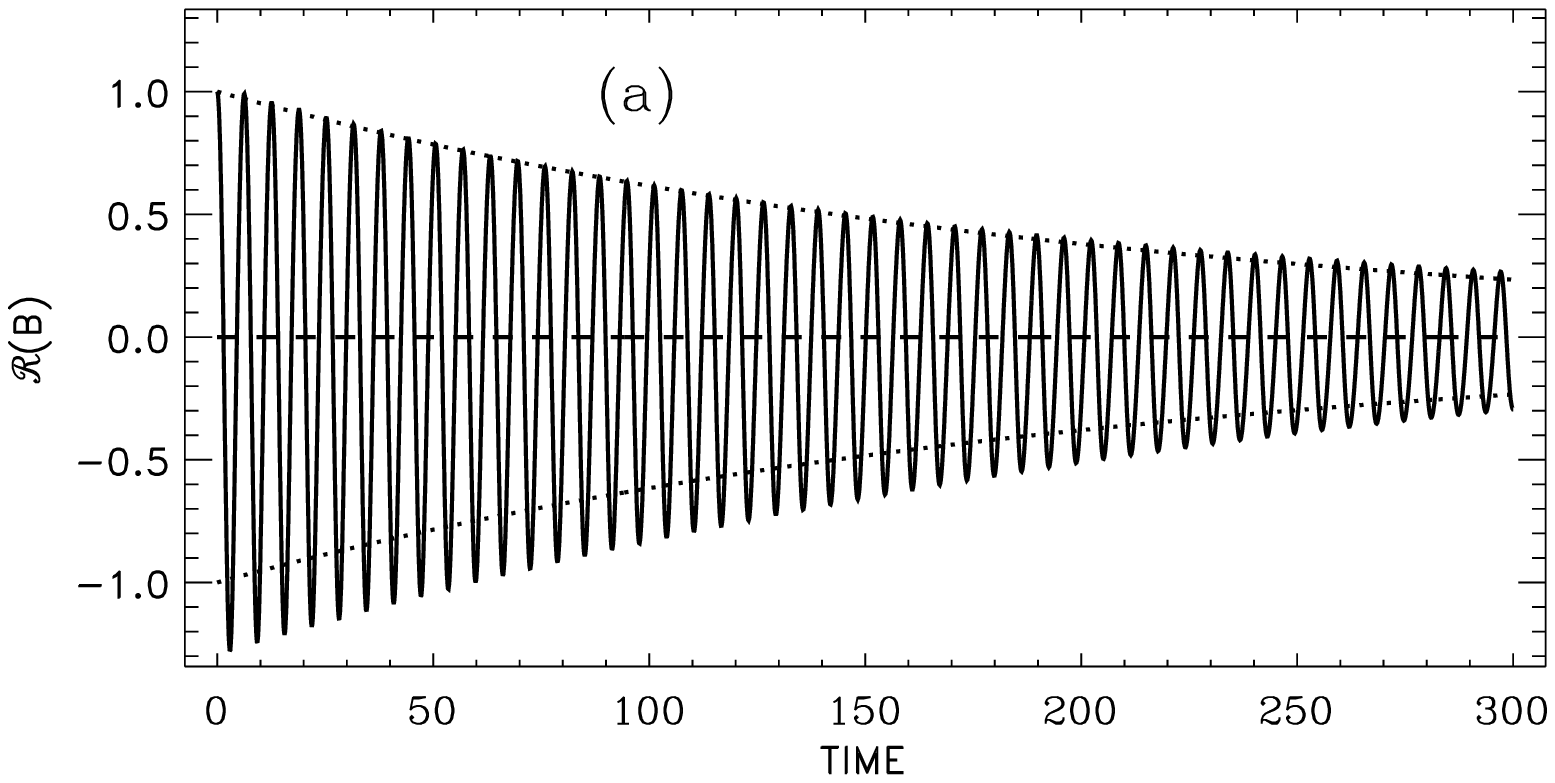}
 \\[0.5 truecm]
 \includegraphics[width=12 cm]{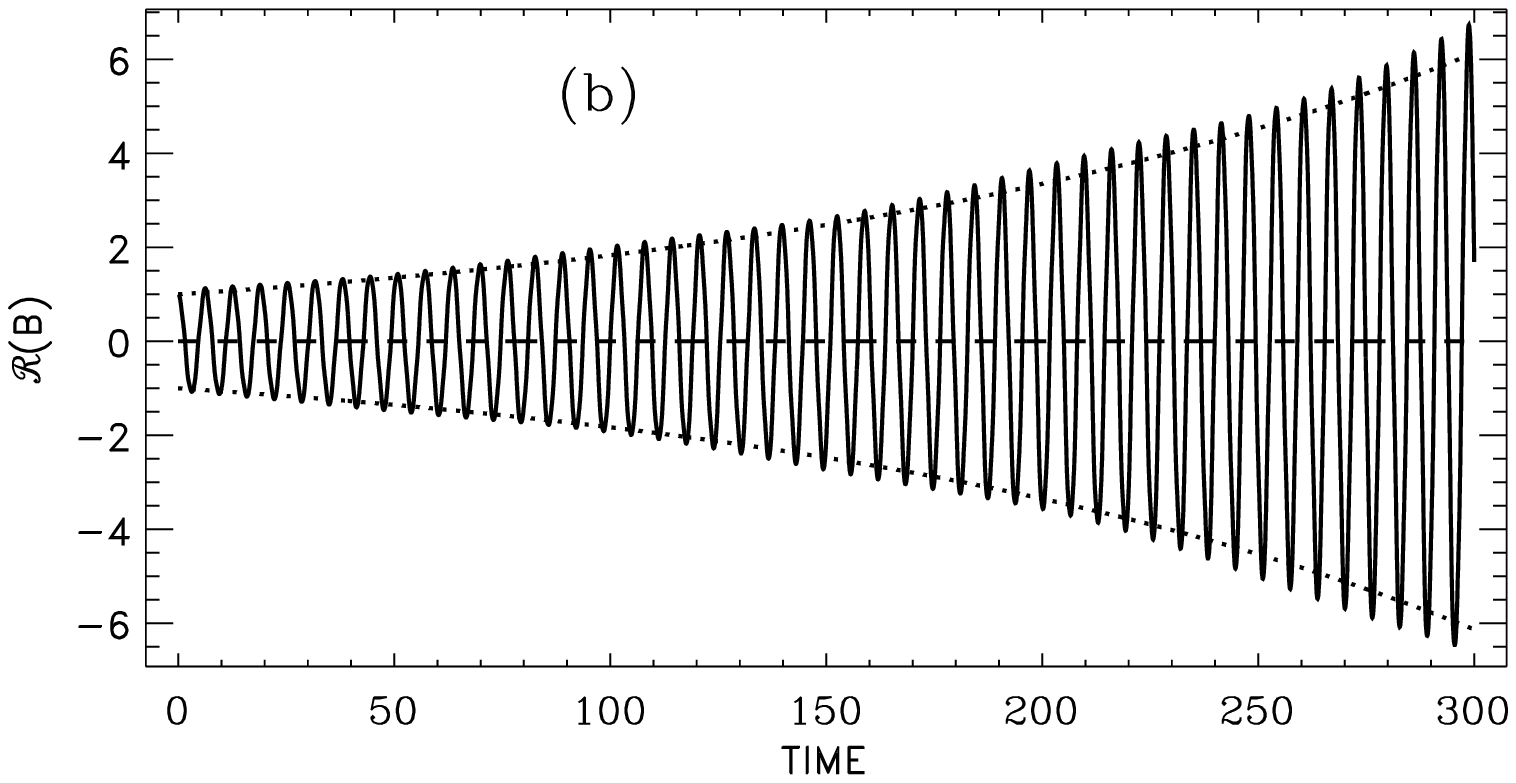}
 \end{center}
 \begin{description}
 \item{\small Fig.~1. Damped oscillations for $\gamma =1$, $\varepsilon = 0.3$ (a) and growing oscillations for $\gamma = 5$, $\varepsilon = 0.7$ (b). The full line shows the numerical solution. The dotted line shows the analytic trends $\pm\mathrm{exp}(st)$ for the oscillation amplitude corresponding to the Eq.\,(\ref{12}).
    }
 \end{description}
\end{figure}

It is of course possible to solve Eq.\,(\ref{10}) numerically to confront the computations with our approximate analytical solution. Figure\,1 gives examples of such a comparison for decaying and growing oscillations. The initial conditions for the computations were $B = 1$ and $\dot{B} = -i$ (for $t = 0$). Equations (\ref{11}) and (\ref{12}) show that the main consequence of the parametric modulation - of the first order in $\varepsilon$ - is the deformation of the oscillations' shape. Variation of the oscillation amplitude is the second-order effect. Part (a) of the Fig.\,1 clearly shows the oscillation asymmetry: negative deflections have larger amplitude.

\begin{figure}[htb]
 \begin{center}
 \includegraphics[width=8.7 cm]{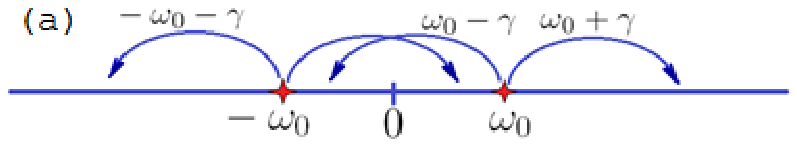}
 \\[0.3 truecm]
 \includegraphics[width=9 cm]{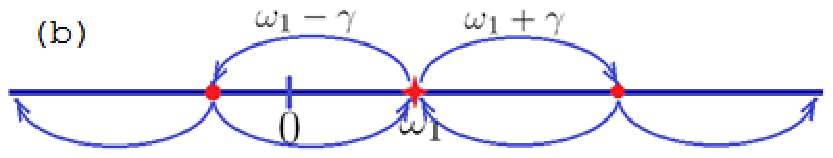}
 \end{center}
 \begin{description}
 \item{\small Fig.~2. (а) Schematic representation of the origin of parametric resonance for oscillations governed by the Mathieu equation (\ref{13}). (b) Non-resonant parametric modulation in a dynamo (cf. explanation in the text).
    }
 \end{description}
\end{figure}

The oscillations grow exponentially for sufficiently large frequency $\gamma$. The growth does not mean a resonance, however. The dependence of the growth rate $s$ on the modulation frequency $\gamma$ of Eq.\,(\ref{12}) is smooth and does not show any selected frequencies characteristic of the resonant phenomena. There is, therefore, no parametric resonance in the dynamo. The reason for this can be seen by comparison of the equation (\ref{9}) with the Mathieu equation for the parametric resonance, which for further discussion is convenient to write as
\begin{equation}
    \ddot{X} + \omega_0^2 X = \varepsilon\sin(\gamma t) X .
    \label{13}
\end{equation}
With parameter oscillations omitted ($\varepsilon = 0$), there are two eigenfrequencies $\omega = \pm\omega_0$. Following the simplest perturbation method (not practical but convenient for qualitative analysis), we can search for the first-order corrections in $\varepsilon$ by substituting the \lq unperturbed' solution $X \propto \mathrm{exp}(\pm i\omega_0 t)$ into the right-hand side of (\ref{13}). This gives a \lq driving force' oscillating with the frequencies $\pm\omega_0 \pm\gamma$ (part (a) in Fig.\,2). For $\gamma = 2\omega_0$, the frequencies $-\omega_0 + \gamma$ and $\omega_0 - \gamma$ meet the eigenfrequencies of the unperturbed system to produce a resonance. This is why the growth rate for parametric resonance is of the first order in $\varepsilon$ (Landau \& Lifshitz 1969). The case with dynamo is different (part (b) in Fig.\,2). There is only one eigenvalue (\ref{7}) for non-damped oscillations. Displacements to the left or to the right by the value $\gamma$ on the frequency axis from this eigenfrequency,  shown by the upper bows in the part (b) of the Figure, do not lead to other eigenfrequencies. The next step of the perturbation technique should be undertaken to get a finite growth rate. This second step gives once again displacements to the left and right on the frequency axis shown by the lower bows in Fig.\,2(b). The second step of the perturbation technique returns to the original eigenfrequency to give a finite growth rate of the order $\varepsilon^2$ (\ref{12}). This happens for any modulation frequency $\gamma$. Parametric modulation in a dynamo is, therefore, \lq not selective' in $\gamma$, i.e. it is not resonant. Parametric resonance is probably possible only for systems with not less than two eigenfrequencies of non-damped oscillations, and a  resonant frequency equal to the difference between these eigenfrequencies.

We turn now to the case of $a(t)$ in equation (\ref{9}) being a random function of time. For this case, the equation has to be solved numerically. Random function with known statistical properties can be realized with the equations
\begin{equation}
    \frac{\mathrm{d}a}{\mathrm{d} t} = -\frac{a}{\tau} + \frac{a_1}{\tau} ,
    \ \ \ \
    \frac{\mathrm{d}a_1}{\mathrm{d} t} = -\frac{a_1}{\tau} + \frac{g}{\tau} ,
    \label{14}
\end{equation}
where $\tau$ is the correlation time and $g$ is a given random forcing. In the computations to follow, $g = \hat{g}\sqrt{4\tau/\Delta t}$, where $\Delta t$ is the numerical time step and $\hat{g}$ is the random number with normal (Gaussian) distribution, the mean value equal to zero and RMS value equal to one. The number is renewed on each time step independent of its preceding value. The function $g$ represents a short-correlated random process but $a(t)$ is a smooth function of time with zero mean. For short $\Delta t \ll \tau$, it represents a stationary (in the statistical sense) random process with the correlation function (Olemskoy \& Kitchatinov 2013)
\begin{equation}
    \langle a(t_0 + t)a(t_0)\rangle =
    \left( \mid t\mid /\tau + 1\right)\mathrm{exp}(-\mid t\mid /\tau) ,
    \label{15}
\end{equation}
where the angular brackets signify averaging over time $t_0$ (see also Rempel 2005).

\begin{figure}[htb]
 \begin{center}
 \includegraphics[width=12 cm]{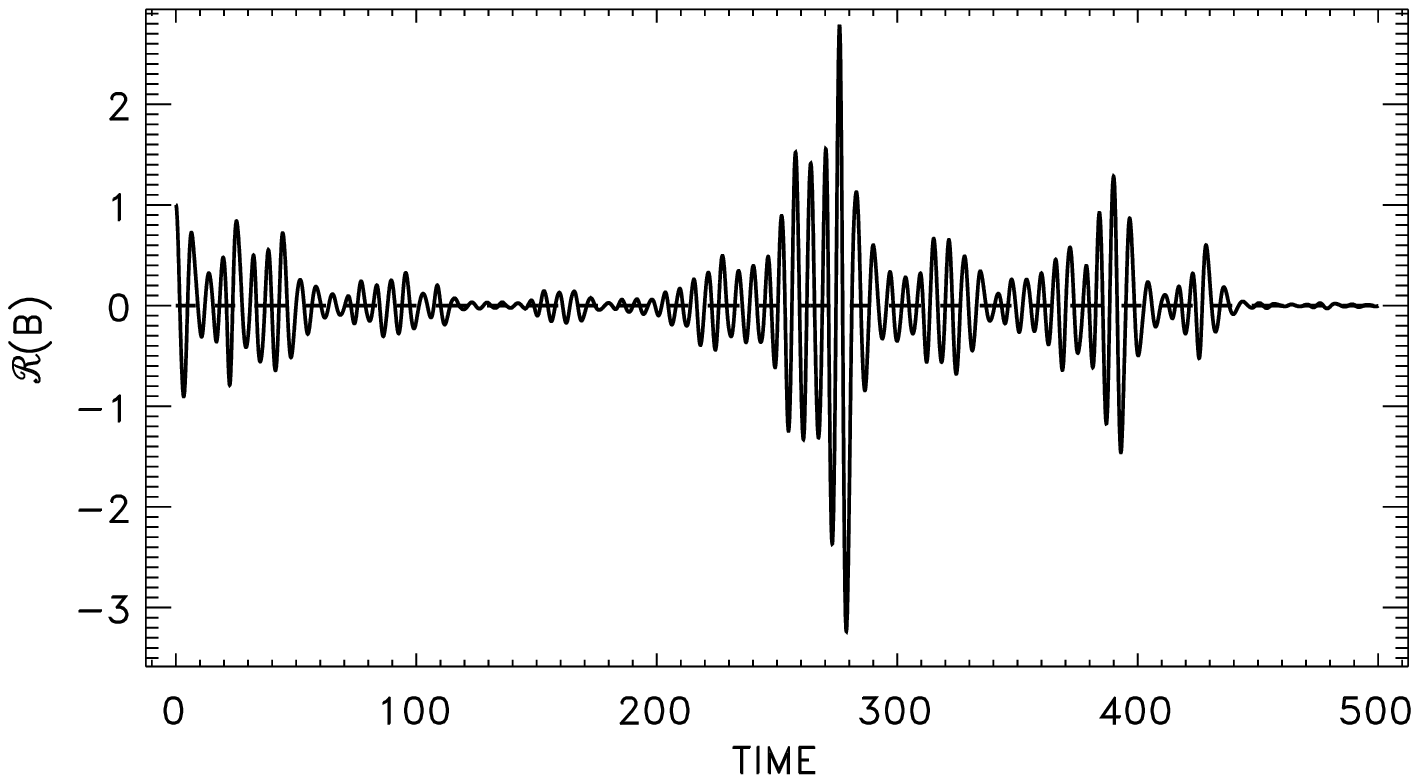}
 \end{center}
 \begin{description}
 \item{\small Fig.~3. An example of linear oscillations in the presence of random
        fluctuations of the $\alpha$-effect. It results from solving jointly the equations (\ref{9}) and (\ref{14}) for $\varepsilon = 0.3$ and $\tau = 1$.
    }
 \end{description}
\end{figure}

Figure~3 shows a characteristic example of the oscillations resulting from a joined solution of Eqs.\,(\ref{9}) and (\ref{14}). Random wandering of the oscillation amplitude away from its initial value is typical of linear oscillation with stochastically varying parameters (Klyatskin 1980). Figure~3 also shows this type of behaviour. The diffusive wandering of the dynamo-wave amplitude is by all probability not realistic. It is stabilized by nonlinear effects, to the account of which we now turn.
 \bll
 \centerline{HYSTERESIS IN NONLINEAR DYNAMOS}
 \bl

Dependence of the coefficients of the equation (\ref{5}) on the magnetic field is now allowed for. The dynamo-wave approximation can still be applied to the nonlinear problem if the coefficients $\hat{\alpha}$ and $\hat{\eta}$ are the functions of the amplitude $\beta = \mid B\mid$ of the wave (see Noyes et al. 1984 for more details). The equation system (\ref{5}) is no longer reducible to one equation similar to Eq.\,(\ref{9}) in the nonlinear problem.

Instead of the Eq.\, (\ref{8}), we now have for the $\hat{\alpha}$-coefficient
\begin{equation}
    \hat{\alpha} = \left( 1 + \varepsilon a(t) \right) \frac{15}{32\beta^4}
    \left( 1 - \frac{4\beta^2}{3 (1 + \beta^2)^2}
    - \frac{1-\beta^2}{\beta}\mathrm{arctg}(\beta )\right) ,
    \label{16}
\end{equation}
where the dependence on the magnetic field is prescribed after the prediction of the quasi-linear theory of the $\alpha$-effect (R\"udiger \& Kitchatinov 1993). Magnetic quenching of the $\alpha$-effect is usually the only nonlinearity in dynamo models. It suffices for stabilizing the filed growth. The $\hat{\alpha}$ value decreases with increasing $\beta$. The effective dynamo number, considered as a function of the magnetic field, reduces to its critical value to stabilize the field growth.

However, not only the $\alpha$-effect but the eddy diffusion as well can depend on the magnetic field.
Quasi-linear theory of turbulent transport coefficients predicts
\begin{equation}
    \hat{\eta} = \frac{3}{8\beta^2}\left( 1
    + \frac{4 + 8\beta^2}{(1 + \beta^2)^2}
    + \frac{\beta^2 -5}{\beta}\mathrm{arctg}(\beta )\right)
    \label{17}
\end{equation}
for the dependence (Kitchatinov et al. 1994). The $\hat{\eta}$, as well as $\hat{\alpha}$, decreases steadily with $\beta$. The effective dynamo number ${\cal D}_\mathrm{eff} = {\cal D}\hat{\alpha}/\hat{\eta}^2$, however, is no longer a monotonic function of the magnetic field. For large $\beta \gg 1$, ${\cal D}_\mathrm{eff} \propto \beta^{-1}$ and the nonlinear saturation of the field growth still takes place. For small $\beta \ll 1$, however, the effective dynamo number increases with the field strength. This leads to the hysteresis phenomenon.

\begin{figure}[htb]
 \begin{center}
 \includegraphics[width=12 cm]{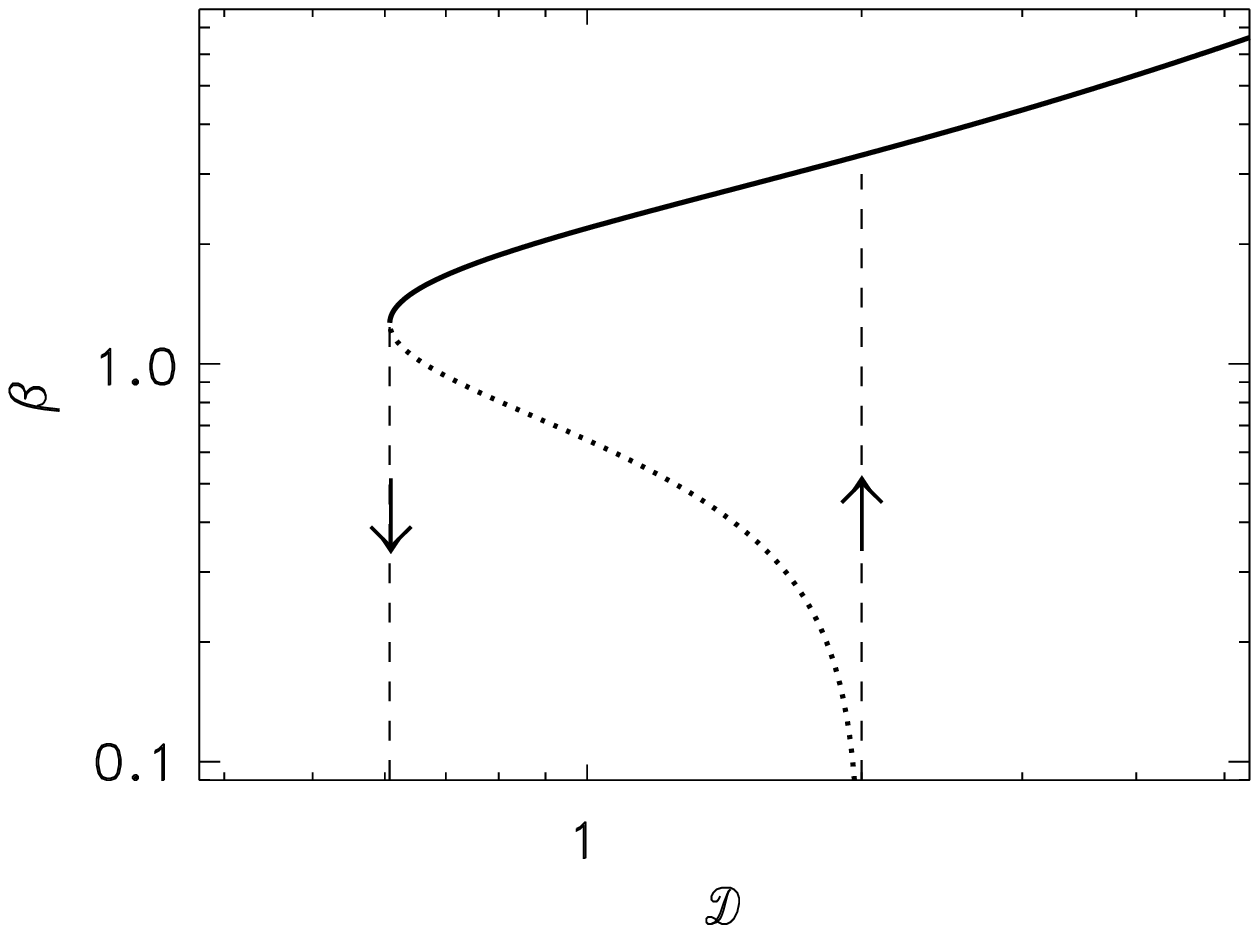}
 \end{center}
 \begin{description}
 \item{\small Fig.~4. Illustration of hysteresis in nonlinear dynamos. In the range of $0.6 < {\cal D} < 2$, the equations (\ref{5}) with the coefficients (\ref{16}) and (\ref{17}) have two stable solutions of i) oscillations with a steady and relatively large amplitude $\beta > 1$, and ii) decaying oscillations of small amplitude. The dotted line shows the watershed between the \lq  attraction regions' of these two solutions.
    }
 \end{description}
\end{figure}

In the absence of $\alpha$-effect fluctuations ($\varepsilon = 0$ in Eq.\,(\ref{16})), there are two stable solutions of Eqs.\,(\ref{5}) in the range of the dynamo numbers $0.6 < {\cal D} < 2$ realised depending on the initial condition. Figure~4 illustrates this situation. Computations for a series of increasing dynamo numbers and a small initial amplitude $\beta$ show decaying fields up to dynamo number ${\cal D} = 2$. When the number $\cal D$ reaches its critical value of ${\cal D}_\mathrm{c} = 2$, the decay is replaced by field growth. This leads to an increase in the effective dynamo-number and the amplitude of the eventually steady oscillations jumps to a finite value of $\beta > 1$. A further increase in dynamo number results in a smooth increase in the amplitude. If the dynamo number is then changed to a decrease and the amplitude of the preceding computation is taken as the initial condition, oscillations of finite amplitude survive for ${\cal D} < 2$ up to ${\cal D} = 0.6$ but decay indefinitely for still smaller ${\cal D} < 0.6$. The equation ${\cal D}_\mathrm{eff}(\beta ) = 2$ for the amplitude of the dynamo-wave has two solutions in the range of $0.6 < {\cal D} < 2$. The solution of the relatively small amplitude is however unstable. This solution is shown by the dotted line in Fig.\,4. This line is the boundary between the \lq attraction regions' of the oscillations with steady and relatively large amplitude (full line) and decaying oscillations. Dynamo waves decay if their initial amplitude is below the dotted line. Above this line, the amplitude grows initially and then stabilizes.

Fluctuations in dynamo parameters can provoke irregular transitions between the two solutions. This possibility attracts attention in connection with the two modes of solar activity inferred by Usoskin et al. (2014) from the content of cosmogenic nuclides in natural archives.
The distribution function of the amplitudes of solar cycles for about 11 thousand years that they obtained has two maxima. Such a dual distribution was interpreted as a manifestation of two distinct modes of solar activity.

\begin{figure}[htb]
 \begin{center}
 \includegraphics[width=12 cm]{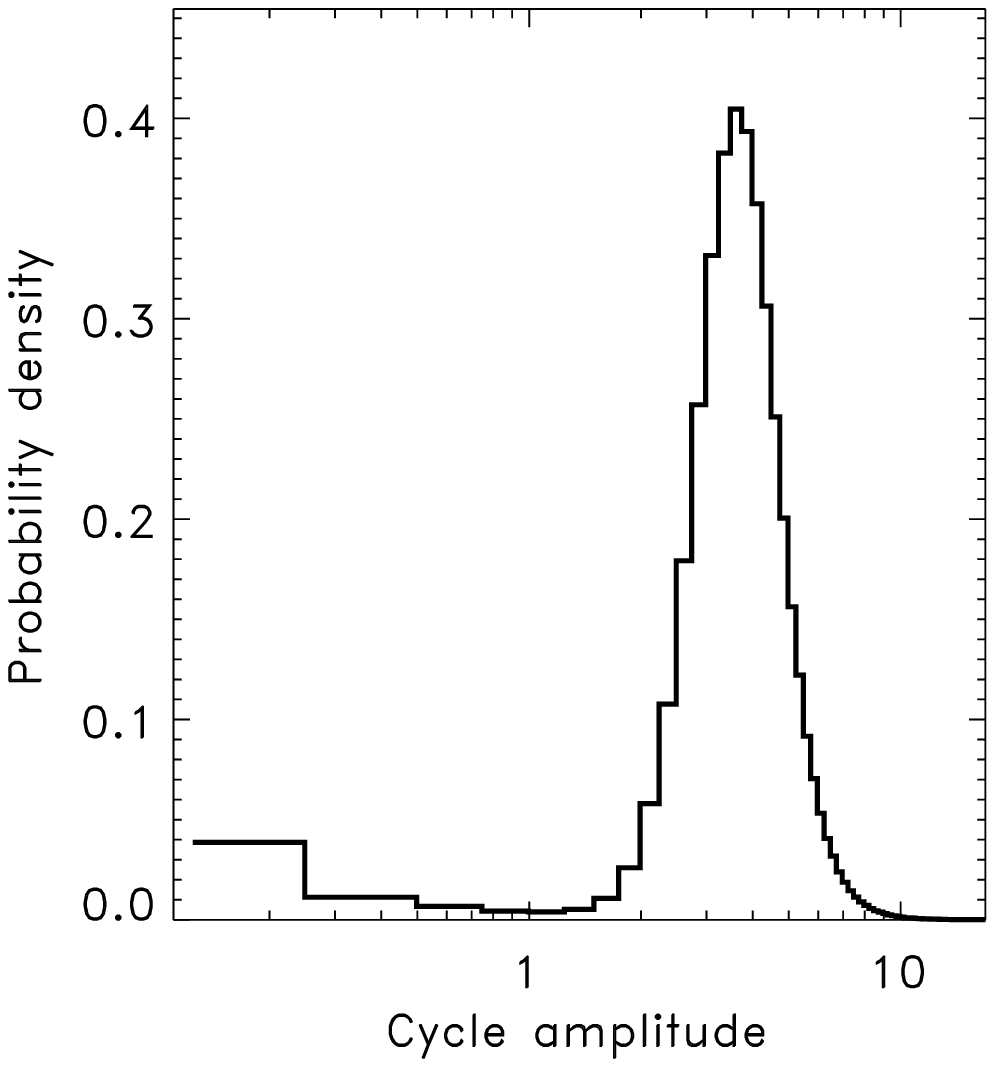}
 \end{center}
 \begin{description}
 \item{\small Fig.~5. Distribution function for the amplitudes of $10^6$ cycles obtained by solving numerically equation (\ref{5}) with allowance for the nonlinearities of Eqs.\,(\ref{16}), (\ref{17}) and fluctuations of the $\alpha$-effect. Computations were performed for ${\cal D} = 2.1$, $\varepsilon = 1$ and $\tau = 1$.
    }
 \end{description}
\end{figure}

Recently Karak et al. (2015) found the dynamo-hysteresis in 3D numerical experiments. Their computations near the critical dynamo number show an irregular intermittency of magnetic cycles of relatively large amplitude with weak-field epochs. However, the time consuming 3D computations did not provide sufficiently vast statistics of magnetic cycles for finding their distribution function. This can be done with the simplified dynamo-equations of this paper. Figure~5 shows the distribution function for one million magnetic cycles computed by solving numerically equation (\ref{5}) with allowance for the nonlinearities of equations (\ref{16}) and (\ref{17}) and the $\alpha$-effect fluctuations of model (\ref{14}). The dual distribution function is qualitatively similar to that found by Usoskin et al. (2014) from historical data on solar activity.

It may be noted that the dynamo-hysteresis and irregular transitions between dynamo modes with different amplitudes of magnetic cycles are not restricted to the special case of the nonlinearities of Eqs.~(\ref{16}) and (\ref{17}). The hysteresis is present whenever the dependence of the ratio $\hat{\alpha}/\hat{\eta}^2$ on the magnetic field has a maximum, e.g., for the simplest model of $\hat{\alpha} = (1 + \beta^3)^{-1}$ and $\hat{\eta} = (1 + \beta )^{-1}$.
 \bll
 \centerline{CONCLUSION}
 \bl

The simplest approximation of the short dynamo-waves of this paper certainly does not allow quantitative reproduction of observations of solar activity. This should be the subject of detailed numerical models. However, the approximation allows analytical solutions, which can be useful in interpreting the results of the numerical computations. Variations of solar activity, including such impressive phenomena as its Grand minima, are related to fluctuations in dynamo parameters. The majority of models for the Grand minima assume the characteristic time of the fluctuations equal to the duration of the (11-year) activity cycle, i.e. to a half of the entire 22-year magnetic cycle that corresponds to the doubled frequency of the magnetic cycle. An enhanced response to fluctuations of such duration as expected for parametric resonance was not found however. Solution of the parametric modulation problem for dynamo-waves shows that parametric resonance is not present in dynamos. Exponential growth or decay of the wave amplitude  depending on the frequency of the parametric variations is possible but a selective sensitivity to special frequencies characteristic of the resonant phenomena does not come out. This is related to peculiarities of the dynamo mechanism (cf. Fig.\,2 and the discussion around). It becomes clear why the Grand minima statistics can be reproduced with either durable (Moss et al. 2008) or rapidly varying (Olemskoy et al. 2013) fluctuations. The fluctuation parameters should be inferred from observations.

The dynamo-wave approximation turns out to be useful also in the problem of the origin of the two modes of solar activity indicated by the new data (Usoskin et al. 2014). The two modes may be explained by the hysteresis in nonlinear dynamos ( Kitchatinov \& Olemskoy 2010; Karak et al. 2015). Hysteresis is also found in the dynamo-wave approximation, which permits its detailed analysis (Fig.\,4). In contrast to more complicated models, computations with this approximation are not numerically demanding. It is therefore possible to accumulate the  statistics of computed magnetic cycles sufficient for constructing a distribution function of the cycles' amplitudes. A dual distribution is obtained qualitatively similar to that found by Usoskin et al. (2014) from radiocarbon data on solar activity.
 \bll

This work was supported by the Russian Foundation for Basic Research (Project \n\ 13-02-00277).
\bll
\centerline{REFERENCES}
\begin{description}
\item Charbonneau,~P.
    2010, \lrsp\ {\bf 7}, 3
\item Choudhuri,~A.\,R., \& Karak,~B.\,B.
    2012, Phys. Rev. Lett. {\bf 109}, 171103
\item Gilman,~P.\,A., \& Dikpati,~M.
    2011, \apj\ {\bf 738}, 108
\item Gilman,~P.\,A., \& Miller,~J.
    1985, \apjs\ {\bf 61}, 585
\item Hoyt,~D.\,V., \& Schatten,~K.\,H.
    1996, \sp\ {\bf 165}, 181
\item Karak,~B.\,B. \& Choudhuri,~A.\,R.
    2013, \raa\ {\bf 13}, 1339
\item Karak,~B.\,B., Kitchatinov,~L.\,L., \& Brandenburg,~A.
    2015, \apj\ accepted, arXiv: 1411.0485
\item Kitchatinov,~L.\,L., \& Olemskoy,~S.\,V.
    2010, \paj\ {\bf 36}, 292
\item Kitchatinov,~L.\,L., \& R\"udiger,~G.
    2004, \aa\ {\bf 424}, 565
\item Kitchatinov,~L.\,L., Pipin,~V.\,V., \& R\"udiger,~G.
    1994, \an\ {315}, 157
\item Klyatskin,~V.\,I.
    1980, {\sl Stochastic Equations and Waves in Randomly-Inhomogeneous Media}
    (Moscow, Nauka Publishers) [in Russian]
\item Landau,~L.\,D., \& Lifshitz,~E.\,M.
    1969, {\sl A Course of Theoretical Physics. Volume 1: Mechanics} (Pergamon Press)
\item Moss,~D.
    1996, \aa\ {\bf 308}, 381
\item Moss,~D., \& Sokoloff,~D.
    2013, \aa\ {\bf 553}, A37
\item Moss,~D., Sokoloff,~D., Usoskin,~I., \& Tutubalin,~V.
    2008, \sp\ {\bf 250}, 221
\item Noyes,~R.\,W., Weiss,~N.\,O., \& Vaughan,~A.\,H.
    1984, \apj\ {\bf 287}, 769
\item Olemskoy,~S.\,V., \& Kitchatinov,~L.\,L.
    2013, \apj\ {\bf 777}, 71
\item Olemskoy,~S.\,V., Choudhury,~A.\,R., \& Kitchatinov,~L.\,L.
    2013, \aj\ {\bf 57}, 458
\item Rempel,~M.
    2005, \apj\ {\bf 631}, 1286
\item Reshetnyak,~M.\,Yu.
    2010, \aj\ {\bf 54}, 1047
\item R\"udiger,~G., \& Kitchatinov,~L.\,L.
    1993, \aa\ {\bf 269}, 581
\item Schmitt,~D., \& R\"udiger,~G.
    1992, \aa\ {\bf 264}, 319
\item Usoskin,~I.\,G., Sokoloff,~D., \& Moss,~D.
    2009, \sp\ {\bf 254}, 345
\item Usoskin,~I.\,G., Hulot,~G., Gallet,~Y. et al.
    2014, \aa\ {\bf 562}, L10
\item Yoshimura,~H.
    1975, \apj\ {\bf 201}, 740
\item Zeldovich,~Ya.\,B., Ruzmaikin,~A.\,A., \& Sokolov,~D.\,D.
    1983, {\sl Magnetic Fields in Astrophysics} (New York: Gordon \& Breach)
\end{description}
\end{document}